\documentclass[twocolumn,aps,prl,showpacs]{revtex4}
\usepackage{graphicx}
\DeclareGraphicsExtensions{ps,eps,epsf}

\newcommand{ \ybco }{\mbox{YBa$_2$Cu$_3$O$_{6+x}$}}
\newcommand{ \ybcoss }{\mbox{YBa$_2$Cu$_3$O$_{6.6}$}}

\newcommand{ \qaf }{\mbox{\textbf{Q}$_{\text{AF}}$}}

\newcommand{ \astar }{\mbox{\textit{a}$^*$}}
\newcommand{ \bstar }{\mbox{\textit{b}$^*$}}
\newcommand{ \cstar }{\mbox{\textit{c}$^*$}}

\newcommand{ \Eres }{\mbox{$E_{res}$}}

\newcommand{ \IC }{\mbox{$\rm \delta $}}

\begin{document}

\title{In-plane anisotropy of spin excitations in the normal and
superconducting states of underdoped \ybco }

\author{ V. Hinkov$^1$, P. Bourges$^2$, S. Pailh\`es$^2$,
Y. Sidis$^2$, A. Ivanov$^3$, C.T. Lin$^1$, D.P. Chen$^1$, and B.
Keimer$^1$ }

\affiliation{$^1$ Max-Planck-Institut f\"ur Festk\"orperforschung,
Heisenbergstr. 1, D-70569 Stuttgart, Germany}

\affiliation{$^2$ Laboratoire L\'{e}on Brillouin, CEA-CNRS,
CE-Saclay, 91191 Gif-sur-Yvette, France}

\affiliation{$^3$ Institut Laue-Langevin, 156X, 38042 Grenoble
cedex 9, France}

\date{\today}

\begin{abstract}
A detailed inelastic neutron scattering study of the in-plane
anisotropy of magnetic excitations in twin-free \ybcoss\ ($T_c =$
61~K) reveals that the spin excitation spectra in the
superconducting and normal states are qualitatively different.
Below $T_c$, the spectrum consists of upward- and
downward-dispersing branches with modest in-plane anisotropy
merging at an energy $\Eres =$ 37.5~meV. In the normal state, the
singularity at \Eres\ disappears, and the spectrum exhibits a
steep dispersion with a strongly anisotropic in-plane geometry.
These data have important implications for models based on static
or dynamic ``stripe" order of spins and charges.
\end{abstract}


\maketitle

A variety of different states with unusual spin, charge, or
current correlations have been invoked to explain the anomalous
normal-state (NS) properties of underdoped copper-oxide
superconductors. Prominent examples are ``striped" states with
spin and charge order extending along one of the principal axes of
the CuO$_2$ square lattice. Neutron scattering experiments yield
detailed information about the microscopic magnetic order and
dynamics and can thus serve as particularly incisive tests of
microscopic models of the copper oxides. Recent neutron scattering
work on $\rm La_{2-x} (Sr,Ba)_{x} CuO_4$ and \ybco\ has uncovered
tantalizing evidence of a ``universal" spin excitation spectrum
independent of materials-specific details
\cite{Pai04,Chr04,Tra04,Hay04,Rez04}. The dispersion surface of
the spin excitations comprises upward- and downward-dispersing
excitation branches merging at the wave vector \qaf\
characterizing antiferromagnetic order in the undoped parent
compounds. The full spectrum thus resembles an ``hour glass" in
energy-momentum space.

Some features of this spectrum agree with calculations for a
specific stripe model, according to which nonmagnetic charge
stripes separate a set of weakly coupled spin ladders in the
copper oxide layers \cite{Uhr04,Voj04,Sei05}. A key prediction of
these models is a pronounced in-plane anisotropy of the magnetic
spectrum. This prediction can, in principle, be tested if a
crystallographically unique in-plane axis pins the stripe
propagation vector. In \ybco\, for instance, the CuO chains along
the $b$-axis define such a direction. In practice, however,
crystallographic twinning (that is, the formation of micron-scale
domains in which $a$- and $b$-axes are interchanged) limits our
ability to extract information about the in-plane anisotropy from
data on the large specimens typically used for neutron scattering
\cite{Ara99,Fon00,Dai01,Sto04,Sto05}. In a recent neutron
scattering study of arrays of small, twin-free \ybco\ single
crystals with x = 0.85 and 0.6, we have provided detailed
information about the in-plane anisotropy of the dynamical spin
susceptibility in the superconducting (SC) state at low energies
\cite{Hin04}. The results disagree with the predictions of the
static stripe model and have stimulated calculations based on
fluctuating stripe arrays \cite{Voj05}.

However, whereas the neutron data were taken in the SC state, none
of the stripe models consider the presence of superconductivity.
In order to provide stringent experimental constraints for
theories based on fluctuating stripes, as well as competing
theories based on anisotropic Fermi liquids and spiral states
\cite{Yam00, Ere05, Zho04, Lin05}, we have carried out experiments
on the spin dynamics in the normal state of twin-free \ybcoss.
Previous work on {\it twinned} samples was unable to resolve the
in-plane anisotropy in the normal state, and the results appeared
to indicate that the NS spin excitation spectrum is simply a
broadened version of the spectrum in the SC state
\cite{Dai99,Sto04}. Our new data now show that both spectra are in
fact {\it qualitatively} different. Specifically, the singularity
at \Eres, which gives rise to the characteristic ``hour glass"
shape of the spectrum, disappears in the NS, and the spectrum
exhibits an unusually steep dispersion with marked in-plane
anisotropy. The low-energy spectral weight is strongly reduced
upon heating above a characteristic temperature $T^*\sim 200$ K.
These results will be discussed in the light of recent theoretical
work.

The experiments were performed on an array of 180 individually
detwinned \ybcoss\ samples with superconducting transition
temperatures (midpoint) of $T_c \approx 61$~K and width $\Delta
T_c \approx 2$~K. The crystals were co-aligned on three Al-plates
with a mosaicity of $<1.2 ^\circ$. The volume of the entire array
was $\sim 450$~mm$^3$, and the twin domain population ratio was
94:6. Measurements were performed at the IN8 spectrometer at the
Institut Laue Langevin (Grenoble, France) and the 2T spectrometer
at the Laboratoire L\'{e}on Brillouin (Saclay, France). Scans
along \astar\ and \bstar\ were carried out under identical
instrumental resolution conditions by working in two different
Brillouin zones. No collimators were used in order to maximize the
neutron flux, and graphite filters extinguished higher order
contamination of the neutron beam.

\begin{figure}[t]
 \includegraphics[width=9.0cm]{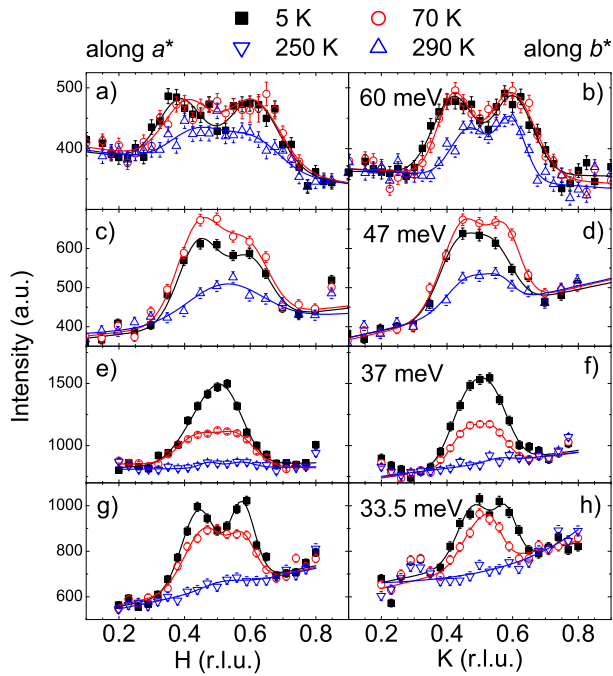}
  \caption{Energy evolution of the in-plane magnetic excitations around
  \qaf\ for three temperatures. Due to kinematic restrictions, data were
  taken in a higher Brillouin zone, along ($H$, -1.5, -1.7) and (1.5, $K$,
  1.7), respectively.
  Close to room temperature the background from multiphonon scattering
  increases. This has been corrected by subtracting a constant
  background at 250~K and 290~K.
  The final wavevector $k_f$ was fixed to $2.66 \mbox{\AA}$ below 38~meV
  and to $4.5 \mbox{\AA}$ above.
  }
  \label{fig1}
\end{figure}

Fig. 1 shows representative raw data in three temperature regimes:
deep inside the SC state at 5~K, just above $T_c$ at 70~K, and at
room temperature. The incommensurability \IC\ (that is, the
deviation of the peak position from $\qaf=(0.5, 0.5)$) and the
spectral weight of the constant-energy cuts are generally
different in the two in-plane directions in reciprocal space, $H$
and $K$, \cite{rlu} and show a complex dependence upon energy and
temperature. A synopsis of the entire data set is shown in Fig. 2
at two temperatures above and below $T_c$. In the following we
will discuss the salient features of the spectrum and its
temperature evolution by referring to Fig. 2 for overall trends,
and to Fig. 1 for more subtle details.

The central result of this paper is the change of the topology of
the dispersion surface from the SC to the normal state. First, we
focus on the spectrum deep in the SC state (Fig. 1, 5~K). Starting
from low excitation energies, \IC\ first decreases with increasing
energy, so that the IC peaks merge at \qaf\ at an energy of
$\Eres=37.5$ meV. For $E > \Eres$, \IC\ increases again, so that
the spectrum forms the ``hour-glass" dispersion (Fig. 2) already
familiar from previous work \cite{Bou00,Hay04,Tra04,Rez04}. In the
NS, however, the singularity at \Eres\ is no longer present, and
the incommensurability is only weakly energy-dependent over a wide
energy range including \Eres\ (Fig. 2c,d). As a consequence, the
NS spectrum no longer shows the ``hour glass" shape that is
characteristic of the magnetic spectrum below $T_c$, and that has
been the subject of much recent debate. This constitutes a
qualitative difference of the spectra in the SC and normal states.

A second major difference between the SC and normal states is
manifested in the in-plane anisotropy. In the SC state, the
dispersion is found to be steeper along \bstar, the direction of
the CuO chains (Fig. 1). The in-plane spectral weight is
moderately anisotropic for $E < \Eres$, that is, the peak
intensity of cuts along \astar\ is higher than that along \bstar.
As pointed out before \cite{Hin04}, the anisotropy decreases with
increasing energy approaching \Eres. A new aspect revealed by the
data in Fig. 1 is that for $E > \Eres$ the spectral weight
anisotropy disappears within the experimental sensitivity. Apart
from the slight difference in \IC\ along the two in-plane
directions, the spectrum is thus fully two-dimensional at high
energies.

\begin{figure*}[t]
 \includegraphics[width=17.3cm]{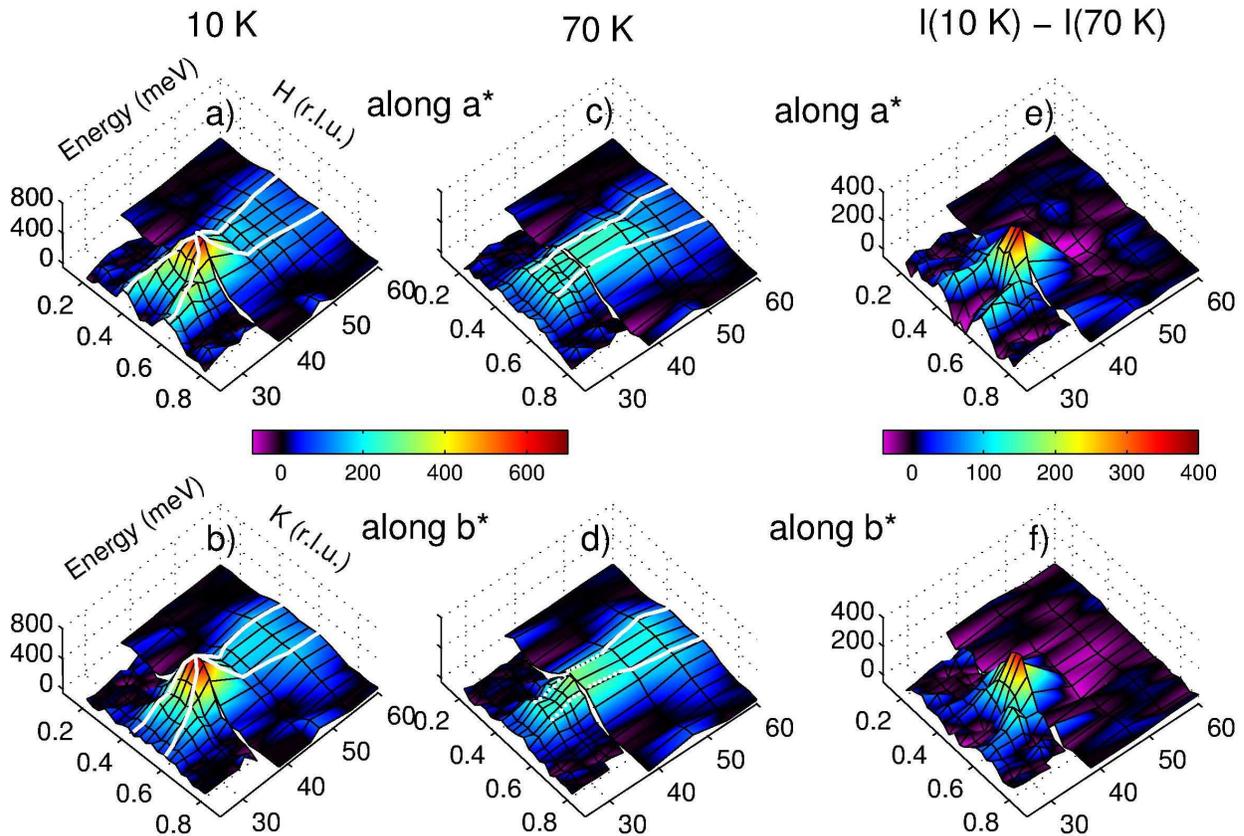}
  \caption{ Color representation of the magnetic intensity. Panels
  a-b show the SC regime, c-d the regime just above $T_c$, and e-f the
  difference between both spectra. The upper and lower rows
  show scans along ($H$, -1.5, -1.7) and (1.5, $K$, 1.7), respectively. In order to obtain a meaningful color
  representation, the intensity at 250 K was subtracted for
  $E <$ 38~meV and the data was corrected for a \textbf{Q}-linear background
  at all energies. The final wave-vector $k_f$ was fixed
  to $2.66 \mbox{\AA}$ below 38~meV and to $4.5 \mbox{\AA}$ above.
  Scans taken at the overlapping energy 38~meV were used to bring both
  energy ranges to the same scale. The white lines connect the fitted peak
  positions of the constant-energy cuts. Dotted lines represent
  upper bounds on the incommensurability.
  }
  \label{fig2}
\end{figure*}

In the NS, the in-plane anisotropy increases strongly for $E <
\Eres$ (70 K profiles in Figs. 1 and 2). Upon heating through
$T_c$, \IC\ and the overall extent of the signal in Q-space
shrinks by about 30\% along \bstar, but only by 10\% along \astar.
The constant-energy profiles become flat-topped, and if these
profiles are fitted to two peaks displaced from \qaf\, the
resulting $a-b$ anisotropy of \IC\ is nearly 40\%, compared to
15\% in the SC state. At energies up to \Eres, the geometry of the
NS spectrum is thus more anisotropic than the one in the SC state.
In contrast, hardly any difference between spectra in the normal
and SC states is discernible at excitation energies significantly
exceeding \Eres. This distinction is highlighted in Fig. 2e-f,
where the difference between the magnetic spectra in the SC and
normal states is shown. The difference signal comprises the
downward-dispersing branch below \Eres, which draws its spectral
weight from a limited range above and a more extended range below
\Eres\ (negative signal in Figs. 2e,f). It is significantly less
anisotropic than the NS spectrum itself (Fig. 2c-d), with respect
to both \IC\ and the spectral weight distribution. Notably, the
difference spectrum is very similar to its analogue in almost
optimally doped $\rm YBa_2 Cu_3 O_{6.85}$, which was shown to
exhibit a nearly circular geometry \cite{Hin04}. This suggests
that the main characteristics of the SC state (such as the gap
anisotropy) are similar at both doping levels.

The spectral rearrangement associated with the formation of the
downward-dispersing branch at $T_c$ results in a sharp upturn of
the intensity at points along this branch (Fig. 3a-b), while at
\qaf\ and 30~meV there is only a broad maximum at $T_c$ (Fig. 3c).
Upon further heating, however, the spectral weight at energies at
and below \Eres\ declines uniformly at all \textbf{Q}-values and
vanishes (to within the experimental sensitivity) at a
characteristic temperature $T^* \approx$ 200~K. This is a further
manifestation of the qualitative difference of the SC and normal
state spectra. $T^*$ is comparable to the temperature below which
various observables exhibit a ``pseudogap" in this doping range
\cite{Tim99}. Corresponding constant-energy cuts show that for $T
> T^*$, the low-energy spectral weight is severely depleted over
the entire Brillouin zone below a characteristic energy $E^* \sim
40$ meV (Fig. 1 e-h, see also Ref. \cite{Dai99}). (Within our
experimental sensitivity, the magnetic spectral weight for $E<E^*$
is indistinguishable from zero. However, it is unlikely that $E^*$
represents a true gap, because \ybcoss\ is metallic at room
temperature, and significant Korringa relaxation has been observed
in NMR experiments.) At energies above $E^*$, in contrast, the
spectral weight is only moderately reduced upon heating above
$T^*$ (Fig. 1 a-d). We can thus distinguish a third temperature
regime above $T^*$, with a magnetic excitation spectrum differing
distinctly from the spectra deep inside the SC state and just
above $T_c$. The intriguing question whether or not the
approximate coincidence of \Eres\ and $E^*$ is accidental should
be addressed by theory.

We now discuss the implications of our observations for stripe
models of the cuprates. On a qualitative level, the large
additional low-energy spectral weight below $T^*$ and the
substantial in-plane NS anisotropy appear compatible with
incipient stripe order. Moreover, recent STM experiments in the NS
of underdoped $\rm Bi_2 Sr_2 Ca Cu_2 O_8$ have revealed low-energy
charge excitations with a ``vertical" dispersion akin to
corresponding features observed in our neutron experiment
\cite{Ver04}. (A related correspondence has recently been pointed
out for the SC state of  $\rm La_{2-x} Sr_{x} CuO_4$, Ref.
\cite{Chr04}.)

\begin{figure}[tl]
 \includegraphics[width=9cm]{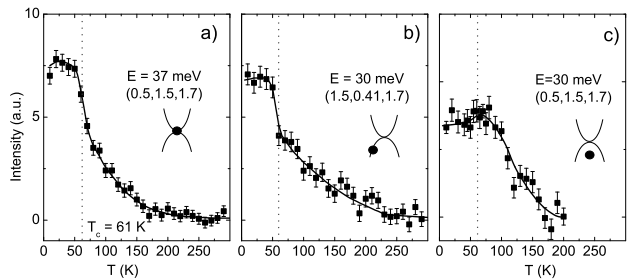}
  \caption{ Temperature dependence of the peak intensity at
  different positions in energy-momentum space indicated in the legend, and sketched
  in reference to the ``hour glass" dispersion in the SC state.
  }
  \label{fig3}
\end{figure}

However, the calculations of the spin excitation spectra of
striped phases reported thus far do not provide a satisfactory
description of our data. In {\it static} stripe models, \Eres\ is
determined by the interaction of spins in adjacent stripes and
represents a saddle point separating a low-energy regime of
anisotropic two-dimensional (2D) excitations from a high-energy
regime with purely 1D excitations \cite{Uhr04,Voj04,Sei05}. At $E
> \Eres$, the neutron cross section is expected to take the form
of streaks in the wave vector direction perpendicular to the
stripes, and only one of the two orthogonal in-plane scans is
expected to show IC peaks, in contradiction to our data in Fig.
1a--d. This definitively rules out static stripe scenarios for
\ybcoss, in agreement with previous conclusions based on
low-energy excitations in untwinned crystals \cite{Hin04} and
high-energy excitations in partially twinned crystals \cite{Sto05}
in the SC state.

The spin excitation spectra of {\it dynamic} stripe phases have
recently been computed numerically \cite{Voj05}. For slowly
fluctuating stripes, the constant-energy cuts exhibit a quasi-2D
intensity distribution, as experimentally observed, but the
spectrum retains its overall ``hour glass" shape. However, since
the influence of superconductivity has thus far not been
considered, predictions of these models should be compared to our
new NS-data, where the ``hour glass" shape is no longer present.
If both SC- and normal-state spectra were found to be amenable to
an interpretation based on fluctuating stripes, our data imply
that the stripe fluctuation rate would have to change dramatically
upon crossing $T_c$. To our knowledge, such a scenario has not
been predicted. Similar considerations apply for other recently
proposed models according to which the ``hour glass" shape of the
magnetic spectrum is a consequence of spiral spin correlations
\cite{Lin05} or other types of modulation \cite{Bat05}.

A natural explanation of the qualitative difference between the
magnetic excitation spectra in the SC and normal states is
provided by models that regard the downward-dispersing branch as
an excitonic mode in the spin-triplet channel below the SC energy
gap \cite{Esc05}. If the downward dispersion is a manifestation of
the d-wave symmetry of the gap, its disappearance in the normal
state is not surprising. Quantitative calculations of the spin
excitation spectrum associated with the triplet exciton have been
carried out mostly in the framework of weak-coupling schemes. It
is questionable whether such models are capable of describing the
strong temperature evolution we have found for $T_c < T < T^*$.
Even in the absence of detailed calculations, however, the
qualitative difference between the spectra above and below $T_c$
appears inconsistent with the proposal \cite{Sto04,Dai99} that the
low-energy spin excitations observed for $T_c < T < T^*$ should be
regarded as an incoherent precursor of the triplet exciton below
$T_c$.

A microscopic explanation of the unusual dispersion and the strong
in-plane anisotropy we have observed in the normal state, as well
as its strong temperature evolution for $T_c < T < T^*$, remains
an important challenge for theoretical work. An interesting
analogy is offered by the spin-1 chain compound $\rm Y_{1-x} Ca_x
Ba Ni O_6$, which exhibits a Haldane gap for $\rm x=0$. For
nonzero density of mobile charge carriers, x, additional spin
excitations with a ``vertical" dispersion develop below the
Haldane gap \cite{Xu00}. The similarity with the
temperature-driven development of spin correlations below the
characteristic energy $E^*$ in \ybcoss\ provides hope that the
doped Haldane chain could serve as a simple model system for a
theoretical description of this behavior.

Acknowledgements: We thank C. Bernhard, G. Khaliullin, D. Manske,
W. Metzner, M. Vojta, and H. Yamase for stimulating discussions.

\bibliographystyle{prsty}

\end{document}